\documentclass{article}
\usepackage[T1]{fontenc} 
\usepackage[utf8]{inputenc} 
\usepackage{amsmath,cite,url}
\usepackage{graphicx}
\usepackage[lbd]{ismir}
\usepackage{booktabs}
\usepackage{color}

\usepackage[firstpage]{draftwatermark}
\definecolor{lightgray}{rgb}{0.9,0.9,0.9}
\definecolor{darkgray}{rgb}{0.4,0.4,0.4}
\SetWatermarkFontSize{12pt}
\SetWatermarkScale{1.1}
\SetWatermarkAngle{90}
\SetWatermarkHorCenter{202mm}
\SetWatermarkVerCenter{170mm}
\SetWatermarkColor{darkgray}
\SetWatermarkText{Late-Breaking / Demo Session Extended Abstract, ISMIR 2025 Conference}

\title{Voices of Civilizations: A Multilingual QA Benchmark for Global Music Understanding}

\multauthor
{Shangda Wu$^{1,\dagger,*}$ \hspace{1cm} Ziya Zhou$^{2,\dagger}$ \hspace{1cm} Yongyi Zang$^{1,\dagger}$}
{\bfseries{Yutong Zheng$^1$ \hspace{1cm} Dafang Liang$^1$ \hspace{1cm} Ruibin Yuan$^2$ \hspace{1cm} Qiuqiang Kong$^3$}\\
 $^1$ Independent Researcher\\
 $^2$ The Hong Kong University of Science and Technology, Hong Kong SAR, China\\
 $^3$ The Chinese University of Hong Kong, Hong Kong SAR, China\\
 $^\dagger$ Core contributors to this work.\\
 $^*$ Corresponding author: \texttt{sanderwood.ir@gmail.com}
}

\def\authorname{S. Wu, Z. Zhou, Y. Zang, Y. Zheng, D. Liang, R. Yuan, and Q. Kong}

\begin{document}

\maketitle
\begin{abstract}
We introduce Voices of Civilizations, the first multilingual QA benchmark for evaluating audio LLMs' cultural comprehension on full-length music recordings. Covering 380 tracks across 38 languages, our automated pipeline yields 1,190 multiple-choice questions through four stages—each followed by manual verification: 1) compiling a representative music list; 2) generating cultural-background documents for each sample in the music list via LLMs; 3) extracting key attributes from those documents; and 4) constructing multiple‐choice questions probing language, region associations, mood, and thematic content. We evaluate models under four conditions and report per‐language accuracy. Our findings demonstrate that even state-of-the-art audio LLMs struggle to capture subtle cultural nuances without rich textual context and exhibit systematic biases in interpreting music from different cultural traditions. The dataset is publicly available on Hugging Face\footnote{https://huggingface.co/datasets/sander-wood/voices-of-civilizations} to foster culturally inclusive music understanding research.

\end{abstract}
\section{Introduction}\label{sec:introduction}
Music is a richly multicultural and multilingual medium, deeply tied to community identity and cultural heritage. However, recent advances in audio LLMs~\cite{DBLP:journals/corr/abs-2311-07919,DBLP:journals/corr/abs-2503-20215,DBLP:journals/corr/abs-2504-18425} have centered largely on high-resource languages and globally dominant genres. Existing benchmarks~\cite{DBLP:conf/ismir/WeckMBQFB24,DBLP:conf/icassp/LiuHSS24,DBLP:conf/icassp/OuyangWZCLL25} further reinforce this bias by excluding traditional, regional, and low-resource musical repertoires. As a result, current models lack exposure to the full spectrum of linguistic and stylistic diversity in music, leaving their performance on non-mainstream content unassessed.

\begin{figure}[t]
    \centering
    \includegraphics[width=0.5\textwidth]{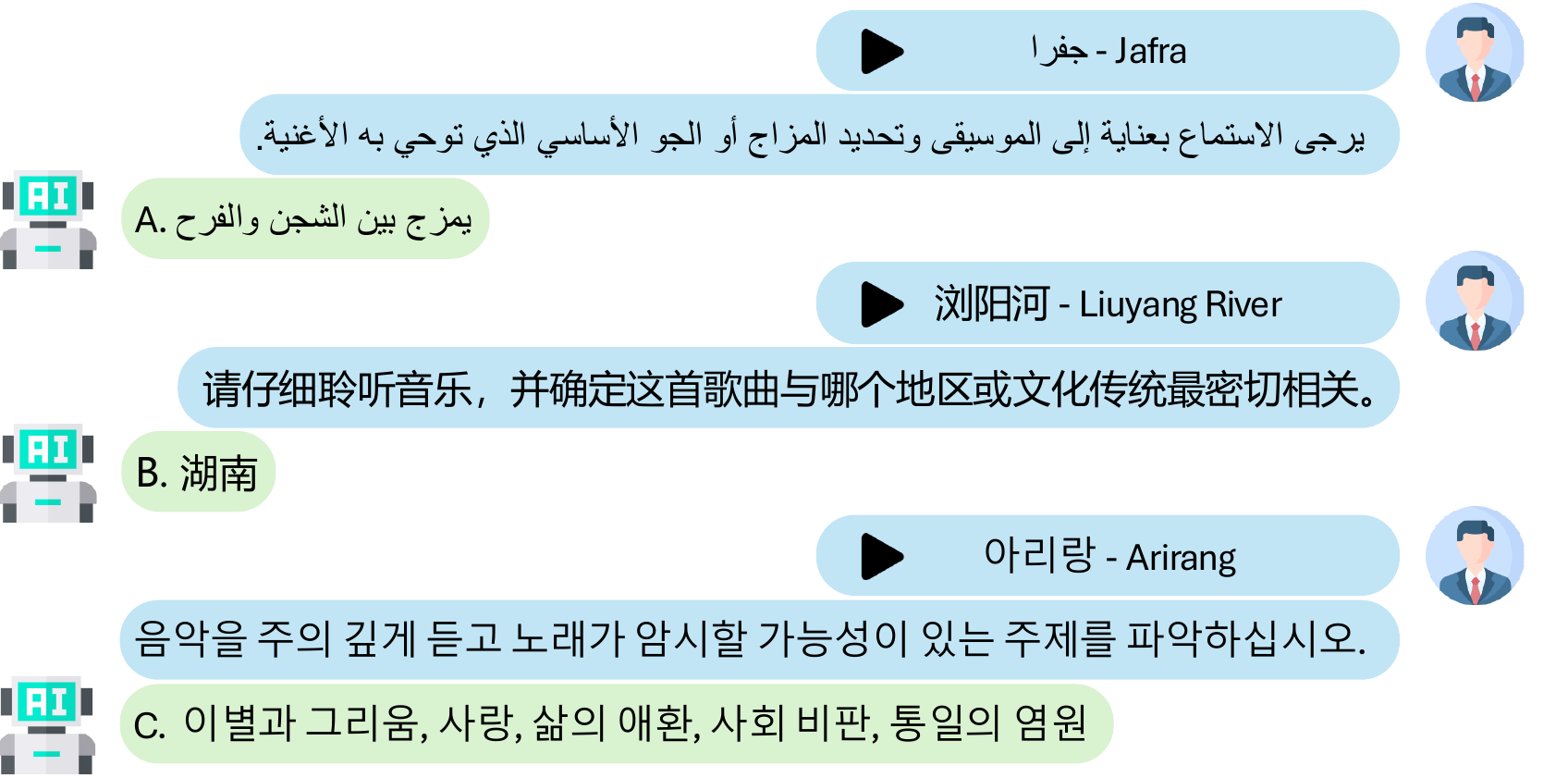}
    \vspace{-2em}
    \caption{Example questions from the Voices of Civilizations benchmark on three folk songs—Arabic "Jafra," Chinese "Liuyang River", and Korean "Arirang."}
    \vspace{-1em}
    \label{fig:demo}
\end{figure}

To address this gap, we present Voices of Civilizations (VoC), the first benchmark to evaluate cultural understanding in audio LLMs using full-length music recordings. It includes 380 tracks in 38 languages and 1,190 multiple-choice questions. Fig.~\ref{fig:demo} gives sample questions from our benchmark. We find that even state-of-the-art models struggle with underrepresented traditions, indicating a need for more culturally inclusive training and evaluation.

\vspace{-1em}
\section{Dataset Construction}\label{sec:dataset}
Inspired by prior work using LLMs to generate music QA benchmarks~\cite{DBLP:conf/ismir/WeckMBQFB24} and collect music-text pairs~\cite{DBLP:journals/corr/abs-2502-10362}, we built VoC using Gemini 2.5 Pro~\cite{comanici2025gemini} in a four-step pipeline:

\textbf{1) Song Selection:}
For each of 38 Gemini-supported languages, we manually selected 10 culturally representative songs from YouTube\footnote{Audio was used solely for non-commercial research under fair use.}, prioritizing traditional styles.

\textbf{2) Context Collection:}
For each song, we collected web documents and used Gemini to generate detailed versions in both the song's native language and English.

\textbf{3) Attribute Extraction:}
Gemini extracted three attributes—region, mood, and theme—from the bilingual documents it generated.

\textbf{4) Question Generation:}
We used Gemini to generate 1,190 multiple-choice questions (language, region, mood, theme). For language, distractors were two random languages plus English; for other types, distractors were drawn from different songs to minimize guessing~\cite{DBLP:journals/corr/abs-2504-00369}.

\begin{figure*}[t]
    \centering
    \includegraphics[width=1.0\textwidth]{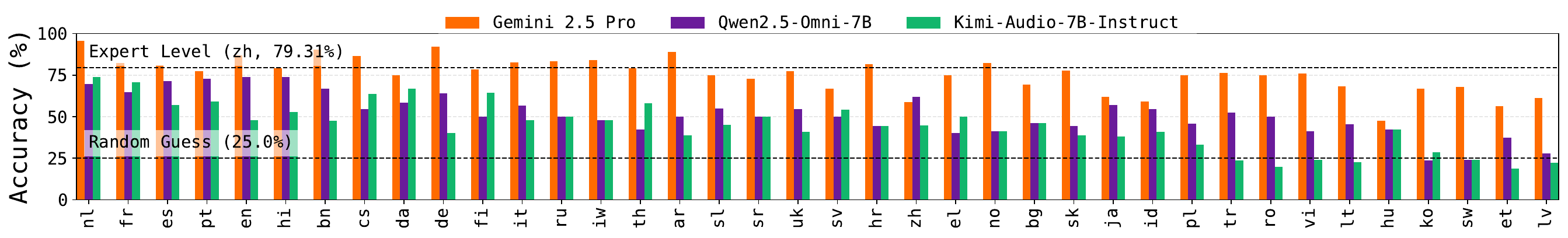}
    \vspace{-2em}
    \caption{Per-language accuracy (\%) of three state-of-the-art audio LLMs on the VoC benchmark using audio input only and focusing on region, mood, and theme questions. We invited a Chinese music teacher to answer 29 questions across 10 Chinese songs in a strictly closed-book setting (no reference or lookup allowed), achieving an accuracy of 79.31\%.}
    \label{fig:language}
\end{figure*}

\begin{table}[t]
\centering
\renewcommand{\arraystretch}{0.95}
\setlength{\tabcolsep}{4pt}
\fontsize{8.5}{9}\selectfont
\caption{Accuracy (\%) of three audio LLMs on four question types in the VoC benchmark under different settings.}
\vspace{0.5em}
\begin{tabular}{lcccc}
\toprule
\textbf{Setting} & \textbf{Language} & \textbf{Region} & \textbf{Mood} & \textbf{Theme} \\
\midrule
\textbf{Noise} & & & & \\
\quad \textit{Gemini 2.5 Pro}   & \textbf{93.42} & \textbf{42.48} & \textbf{40.15} & \textbf{47.50} \\
\quad \textit{Qwen2.5-Omni-7B}     & 51.05 & 26.11 & 23.48 & 31.25 \\
\quad \textit{Kimi-Audio-7B-Instruct} & 40.26 & 23.11 & 23.45 & 24.06 \\
\midrule
\textbf{Audio (Eng QA)} & & & & \\
\quad \textit{Gemini 2.5 Pro}   & \textbf{99.74} & \textbf{73.01} & \textbf{62.50} & \textbf{85.00} \\
\quad \textit{Qwen2.5-Omni-7B}     & 86.32 & 46.02 & 51.89 & 56.25 \\
\quad \textit{Kimi-Audio-7B-Instruct} & 85.79 & 40.91 & 41.15 & 48.75 \\
\midrule
\textbf{Audio} & & & & \\
\quad \textit{Gemini 2.5 Pro}   & \textbf{100.00} & \textbf{75.22} & \textbf{62.12} & \textbf{87.19} \\
\quad \textit{Qwen2.5-Omni-7B}     & 89.47 & 44.25 & 50.38 & 58.13 \\
\quad \textit{Kimi-Audio-7B-Instruct} & 85.26 & 42.05 & 41.15 & 50.62 \\
\midrule
\textbf{Audio + Doc} & & & & \\
\quad \textit{Gemini 2.5 Pro}   & \textbf{100.00} & \textbf{99.12} & 89.39 & \textbf{98.12} \\
\quad \textit{Qwen2.5-Omni-7B}     & 97.37 & 97.35 & 83.71 & 94.69 \\
\quad \textit{Kimi-Audio-7B-Instruct} & 93.42 & 81.06 & \textbf{93.36} & 92.81 \\
\bottomrule
\end{tabular}
\label{tab:voc_settings}
\end{table}

\section{Experiments}\label{sec:experiments}
We evaluate three state-of-the-art audio LLMs—Gemini 2.5 Pro~\cite{comanici2025gemini}, Qwen2.5-Omni-7B~\cite{DBLP:journals/corr/abs-2503-20215}, and Kimi-Audio-7B-Instruct~\cite{DBLP:journals/corr/abs-2504-18425}—on the VoC benchmark under four settings. Unless otherwise noted, all questions are presented in their original language:

\begin{itemize}
    \item \textbf{Noise:} All models receive the same 10-second Gaussian noise instead of music, providing a lower bound to assess reliance on audio content.
    \item \textbf{Audio (Eng QA):} Full music audio paired with English questions, measuring baseline comprehension independent of multilingual grounding.
    \item \textbf{Audio:} Full music audio paired with questions in the same language as the song, testing cross-lingual understanding from audio alone.
    \item \textbf{Audio + Doc:} Full music audio paired with its background document in the corresponding language, assessing the benefit of additional textual context.
\end{itemize}

As Gemini contributed to VoC construction, the evaluation favors it in that the documents match its generation style, the languages are within its supported set, and the distractors (except in language questions) were selected by the model itself. However, the audio recordings were independently curated, without Gemini's involvement. Therefore, in audio-only settings—\textbf{Audio (Eng QA)} and \textbf{Audio}—the model must rely on its own audio understanding.

Table~\ref{tab:voc_settings} reports accuracy by question type across conditions. In the \textbf{Noise} setting, all models perform above chance on \textbf{Language} questions, apparently guessing based on the language of the question. Qwen and Kimi score near chance on non-language questions, confirming that distractors drawn from other songs' attributes suppress guessing. Gemini performs notably better, likely due to stronger reasoning and alignment with distractors it selected. In both \textbf{Audio} and \textbf{Audio (Eng QA)} settings, all models achieve high accuracy on \textbf{Language} questions (>85\%), suggesting that language identification is relatively easy, as it primarily relies on simple phonetic cues at the speech level. In contrast, performance on \textbf{Region}, \textbf{Mood}, and \textbf{Theme} remains low with audio alone, indicating limited ability to extract cultural meaning from sound. Switching the question language from English to the song's native language has little effect, suggesting multilingual question comprehension is not a major limitation. Accuracy improves substantially only when background documents are provided, with all models—especially Gemini—approaching near-perfect scores, highlighting their reliance on textual context over audio-based cultural reasoning.

To assess model robustness across cultural space, Fig.~\ref{fig:language} reports per-language accuracy (\textbf{Audio}), aggregated over all question types except language identification. Performance is highly non-uniform: high-resource languages tend to score higher, while under-represented and low-resource traditions show sharp declines. This long-tail pattern appears across models, underscoring the need for culturally and linguistically diverse training data.

\section{Conclusions}\label{sec:conclusions}
We present Voices of Civilizations (VoC), the first multilingual benchmark for evaluating cultural understanding in audio LLMs through full-length music. Featuring 1,190 questions across 38 languages, VoC reveals substantial performance gaps—particularly on music from under-represented languages and cultural traditions.

Designed as a proof of concept, VoC focuses on cultural attributes, as low-level musical features (e.g., key, meter, tempo) remain unreliable to extract. Its scope is intentionally narrow: few songs per language, no instrumentals, a single question style, and omits how musical works evolve and spread. We release VoC to invite the community to broaden its coverage—so that music from marginalized cultures gains greater visibility in evaluation and greater influence in the training of future audio LLMs.

\bibliography{ISMIRtemplate}

%
%
%
%
%

\end{document}